\documentclass[a4paper,twocolumn]{esapub} 

\usepackage{graphicx}
\usepackage{times}
\usepackage{natbib}

\def\degr{\hbox{$^\circ$}}
\def\arcmin{\hbox{$^\prime$}}

\begin{document}

\pagestyle{empty}

\title{\large \bf Initial results from SECIS observations of the 2001 eclipse}

\author{\bf A.C.\ Katsiyannis$^{(1)}$, R.T.J.\ M$^{\mathrm{c}}\!$Ateer
                  $^{(1)}$, D.R.\ Williams$^{(2,1)}$, P.T.\
                  Gallagher$^{(3)}$, F.P.\ Keenan$^{(1)}$}

\affil{\it $^{(1)}$Department of Pure and Applied Physics, 
           Queen's University Belfast, Belfast, BT7 1NN, U.K.,
           a.katsiyannis@qub.ac.uk and j.mcateer@qub.ac.uk and
           f.keenan@qub.ac.uk}

\affil{\it $^{(2)}$Mullard Space Science Laboratory, University College
           London, Holmbury St. Mary, Dorking, Surrey, RH5 6NT, U.K.,
           drw@mssl.ucl.ac.uk}

\affil{\it $^{(3)}$ L-3 Communications GSI, NASA Goddard Space Flight 
           Center, Greenbelt, MD 20771, U.S.A., 
           Peter.T.Gallagher@hessi.gsfc.nasa.gov}

\maketitle

\abstract{SECIS observations of the June 2001 total solar eclipse 
were taken using an Fe~{\sc xiv} 5303 {\AA} filter. Existing software
was modified and new code was developed for the reduction and analysis
of these data. The observations, data reduction, study of the
atmospheric and instrumental effects, together with some preliminary
results are discussed. Emphasis is given to the techniques used for
the automated alignment of the 8000 images, the application of the \`a
Trous algorithm for noise filtering and the software developed for the
automated detection of intensity oscillations using wavelet
analysis. In line with findings from the 1999 SECIS total eclipse
observations, intensity oscillations with periods in the range of
20-30~s, both inside and just outside coronal loops are also
presented.}

\section{\bf Introduction}

Coronal loops are known to be subject to different types of
magnetohydrodynamic (MHD) oscillations. Review papers by Aschwanden et
al. [1], Nakariakov [2], Roberts[3] and others provide an overview of
the theoretical and observational progress on the detection of such
oscillations in solar coronal loops. It is believed that the study of
those events may provide us with more information about the physical
characteristics of the corona (a subject area called coronal
seismology) and help us investigate the feasibility of the coronal
oscillations as a corona heating mechanism.

It has been suggested that MHD waves can be divided into two main
categories. Magnetoacoustic waves, consisting of density, pressure and
temperature perturbations (which in turn are divided into slow and
fast modes) and incompressible Alfv\'en waves (which are also divided
into those with movements perpendicular to the magnetic field and the
torsional oscillations). Roberts et al.\ [4] provide a detailed
theoretical study, using reasonable approximations for a low-$\beta$
plasma, of the fast-mode magnetoacoustic perturbations and predict
that fast magnetoacoustic (sausage-mode) oscillations with periodicity
of $\sim$1 Hz may be excited in coronal loops. The same authors also
predict how the signature of a wave train, created by a pulsation and
propagating to a distance {\it h} from it source, could provide us
with information of the physical parameters of the loop. More recently
Nakariakov et al. [5] have confirmed the analytical work of Roberts et
al.\ [4] on the propagation of wave trains, by numerically modelling a
perturbing pulse and simulating the time series as developed at a
distance {\it h} along the loop.

The coronal heating problem is addressed by a large number of authors
with two main mechanisms proposed as the most likely explanation (see
review article by Priest \& Schrijver [6]). One of these is that a
large number of magnetic reconnections causing current dissipation
could result in micro- or nano-flare activity on a regular bases (see
Parker [7] more further discussion). The other suggestion is the
damping of MHD waves, caused by ion viscosity and electrical
resistivity (first introduced by Hollweg [8]).

Following a long sequence of attempts to detect coronal oscillations
using total solar eclipse observations (Koutchmy et al.\ [9],
Pasachoff \& Landman [10], Singh et al.\ [11], Cowsik et al.\ [12],
Pasachoff et al.\ [13]), the Solar Eclipse Coronal Imaging System
(SECIS) observed the 1999 total solar eclipse using an Fe~{\sc xiv}
5303 {\AA} filter. Phillips et al.\ [14] developed SECIS and used the
instrument to observe the total solar eclipses of 1999 and 2001. They
succeeded in detecting several periodicities in the range of 4-7 s
(Williams et al.\ [15], Williams et al.\ [16], Katsiyannis et al.\
[17]) in the 1999 data set. A propagating wave train was detected by
[16] with a phase-speed of $\sim$2100~km~s$^{-1}$, reinforcing the
identification of this perturbation as a fast-mode MHD
wave. Considering the physical parameters of the loop (physical
dimensions, density, etc) and making reasonable assumptions about the
strength of the magnetic field, it was found that the frequencies
detected are well within the range of predicted values in [4]. Work on
different loops in the same active region [17] failed to detect a wave
train but found intensity oscillations just outside these loops. These
perturbations were in the same frequency range as those detected by
[16] and with similar amplitudes.

In this paper we discuss observations taken during the June 2001 total
solar eclipse, the reduction and data analysis techniques developed to
process these data and some preliminary detections of oscillations in
the lower solar coronal.


\section{\bf Observations}

For a detail description of the instrument and details of the August
1999 observations, see [14]. Since the previous eclipse, several
alterations were made to the system, to improve the performance of the
instrument.

\begin{itemize}

\item A broader, Fe~{\sc xiv} 5303 {\AA} filter was used. While both 
filters were centred on the 5303 {\AA} line, the previous filter had
full-width-half-maximum (FWHM) of 2 {\AA} while the new filter has a
FWHM of 5 {\AA}.

\item A metallic cover was produced to seal the optical elements of 
SECIS from scattered light. The area covered extended from the back of
the Schmidt telescope to the charge-coupled device (CCD) cameras (see
[14] for more details on the instrument layout).

\item The driving mechanism of the heliostat was covered to protect 
against particles of dust or mud. This was considered important as any
contamination could cause a temporarily change of the rotation speed
of the heliostat.

\item Cooling fans were installed on the CCD cameras to minimize high
temperature effects.

\end{itemize}

Similar procedures to the August 1999 observations were followed for
the 21 June 2001 total solar eclipse. The instrument was located on
the roof of the physics department of the University of Zambia, in
Lusaka, Zambia (Latitude: 15$\degr$ 20$\arcmin$ South; Longitude:
28$\degr$ 14$\arcmin$ East). The instrument was transported to the
location in parts and assembled on the spot a few days before the
eclipse. The heliostat was aligned to the site's local meridian by
using the standard gnomon technique. The optical components were
aligned using a small pocket laser and the instrument was focused
using very distant objects. On the day of the eclipse weather
conditions were very good, with practically no wind nor clouds. The
cooling fans were switched on early in the morning of the eclipse, but
were switched off minutes before totality to avoid causing vibrations
to the instrument.

The CCD detectors each have 512 $\times$ 512 pixel$^{2}$, which
combined with the instrument's optics provide us with a resolution of
$\sim$ 4 arcsec~pixel$^{-1}$ (see [15] and references therein) and an
observable area of $\sim$ 34$\times$34 arcmin$^{2}$. As edge effects
of the CCD prevent us from using all 512 $\times$ 512 elements (only
an area of $\sim$ 400 $\times$ 300 pixel$^{2}$ is considered to be
free of edge effects) and in line with the practice we followed during
the 1999 total solar eclipse, we decided not to observe the whole
disk, but only the North-East limb. The choice of location was based
on the appearance of NOAA Active Region 9513 on the limb of the disk
in the same general area the previous day. Coordinated Solar and
Heliospheric Observatory (SoHO) observations of the same area were
taken during totality in order to provide us with the ability to
determine the physical characteristics of any coronal regions detected
with intensity oscillations. The SECIS instrument itself obtained 8000
images at a rate of 39 frames per second covering the whole duration
of totality.

The next morning sky flats and dark frames were taken. For the
sky flats we used the same exposure time as with the eclipse
observations, while for the dark frames we covered the CCD cameras
completely with a black cloth, closed the aperture of the lenses to
f/22 and set the exposure time equal to that used during eclipse.


\section{\bf Data Reduction and Analysis}

As SECIS is a project-specific instrument, no data reduction software
is widely available and with the exception of some subroutines written
for the August 1999 observations, the subroutines used were developed
from scratch. As some of the procedures applied are not commonly used
in astronomical software, a detailed description follows below.

\subsection{Image Alignment}

The first part of the data reduction included dark current and sky
flat subtraction performed as is normal for most astronomical
observations. The next part of the reduction was less common in
astronomical data reduction as the images had to be aligned with an
accuracy of a fraction of a pixel in order to achieve the accuracy
needed for the study of intensity oscillations. One of the
difficulties was that, unlike the 1999 observations, no prominence was
present on the limb of the sun at the time of the eclipse and hence
the previous alignment procedure (see [15] for a detailed description)
had to be modified. Although the first part of the alignment of the
2001 data set was very similar to that of the previous observations,
the whole procedure will be described for completeness.

A first order approximation of the position of the lunar disk was
achieved by using the Sobel filter to calculate the edge of the Moon
on each individual frame. Then, assuming a constant lunar radius, a
fit of the disk was produced by using the least square method and
rejecting all points lying outside more than 3$\sigma$ outside the
best fit. The next step was to assume the motion of the Moon with
respect to the Sun was of constant angular velocity and thus aligning
the images with an accuracy of $\sim$ 1 pixel.

For further accuracy in alignment, a second stage was introduced to
the process. An area of a reference image containing featureless parts
of the corona and with a sharp transition to the moon disk was
identified and expanded by a factor of 20 along each axis. For the
8000 individual images, the same area was expanded by a factor of 20
and those areas were cross-correlated with the reference image. Then
for every frame the pixel shift that gave the best correlation was
divided by 20 and stored as the shift of that frame. The whole,
unexpanded images were shifted by those values using bilinear
interpolation for non-integer shifts. This procedure thus aligned the
images to an accuracy of 0.05 pixel.

\subsection{Wavelet analysis} 

A continuous wavelet transformation was used to analyse the
observations described for two main reasons:

Firstly, wavelet algorithms have gradually started to replace the
classical Fourier analysis as it provides us with localised temporal
information.  Coronal oscillations are not necessary expected to last
longer than a few periods, therefore high frequency oscillations with
periodicities of a few seconds may not last longer than a few tens of
seconds, i.e., much shorter than the $\sim$ 3 min of the SECIS 2001
observations. Fourier analysis has, by nature, a limited ability to
detect oscillations lasting a fraction of the time series, making the
wavelet analysis the method of choice. Torrence \& Compo [18] have
described in detail this algorithm and provided a discussion of the
benefits of it's application on different scientific fields.

Secondly, the increased popularity of the wavelet transformation (for
example, Gallagher et al.\ [19]; Ireland et al.\ [20]; Banerjee et
al.\ [21]) and its consistent use throughout the analysis of the 1999
SECIS observations ([15], [16], [17]), make it an obvious choice as it
allows us to make direct comparisons with recent work in this subject
area.

A Morlet wavelet was used for the analysis of our data, with waveform

\begin{equation}
\psi(\eta)=\pi^{-1/4}\exp(i\omega_0\eta)\exp(\frac{-\eta^{2}}{2}),
\label{morlet}
\end{equation}

\noindent where $\eta = t/s$ is the dimensionless time parameter, t
is the time, s the scale of the wavelet (i.e. its duration), $\omega_0
= s\omega$ is the dimensionless frequency parameter, and $\pi^{-1/4}$
is a normalization term (see [18]).

\begin{figure}
\centering
\includegraphics[bb=20 20 575 470, width=8cm]{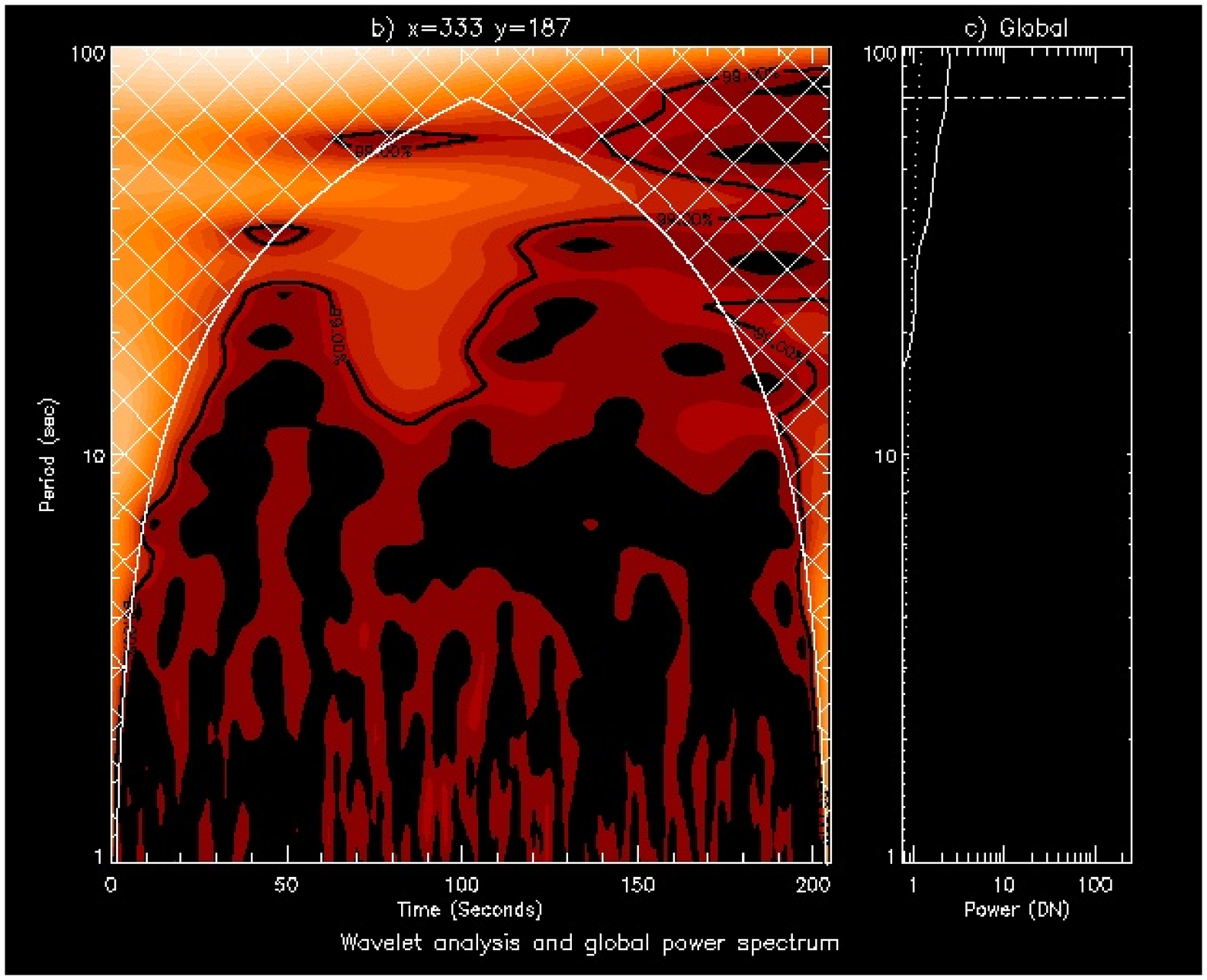}
\caption{Wavelet transform analysis of point x=333, y=187 of the 
    aligned data set. (a) contains the wavelet transform of the time
    series and (b) the global wavelet spectrum. The contours in panel
    (b) highlight the areas where the detected power is at the 99\%
    confidence level and the hatched area, the cone-of-influence
    (COI). }
\label{wavelet}
\end{figure}

The results of the wavelet analysis as described above applied to the
x=333, y=187 pixel of the aligned 2001 data set can be seen in Figure
1. It is divided into two areas with panel (a) showing the power
density wavelet transform with the lighter areas representing the
higher values. The hatched region marks the cone-of-influence (COI)
and represents the areas that suffer from edge effects. Everything
inside the COI is discarded for the purposes of this work. For further
discussion on the edge effects introduced by a finite time-series see
[18] and references therein. The contours of panel (a) surround the
area where the detected power exceeds the 99\% confidence level.

Panel (b) contains the global wavelet spectrum, which is the wavelet
analogue of the standard Fourier transform. It is produced by summing
the power density wavelet transform over the whole time series, while
the dotted line running along the period axis is the global
significance level (again summed over time) at the same value (99\%)
as the contours in panel (a). The horizontal dot-dashed line near the
top of panel (b) marks the bottom of the COI and all detections below
this frequency should be discarded.

\subsection{Noise Reduction}

SECIS has been designed as a highly portable and autonomous
astronomical instrument, made to be able to observe in any location
with a minimum of infrastructure requirements. As such, one of the
most significant limitations of the instrument is that the 200 mm
aperture of the Meade Schmidt-Cassegrain $f/10$ telescope (see [14]
for more details), in conjunction with the ultra-fast sampling rate of
the cameras, results in a low signal-to-noise ratio (S/N).

Various noise reduction possibilities were investigated and the \`a
Trous filtering was chosen as the optimal method. This is a standard
mathematical procedure developed for any time-series affected by
either Gaussian or Poison noise, to be analysed by wavelet
transformations. It is commonly applied to astronomical image and data
analysis and readily available in literature. Starck \& Murtagh [22]
provide a detailed description of the method and its applications. The
non-entropy-based variety of the \`a Trous method was chosen for
simplicity.

The sequence of 8000 frames taken during the eclipse is, in effect, a
three-dimensional array where the x, y axes are the CCD's two
dimensions and the z axis is the image number. This array can be
divided into vectors (or time series) of the same x, y location
extending through the whole range of z. The 1-d \`a Trous algorithm
can then be applied to each such time series separately.

\begin{figure}
\centering
\includegraphics[bb=20 20 575 477, width=8cm]{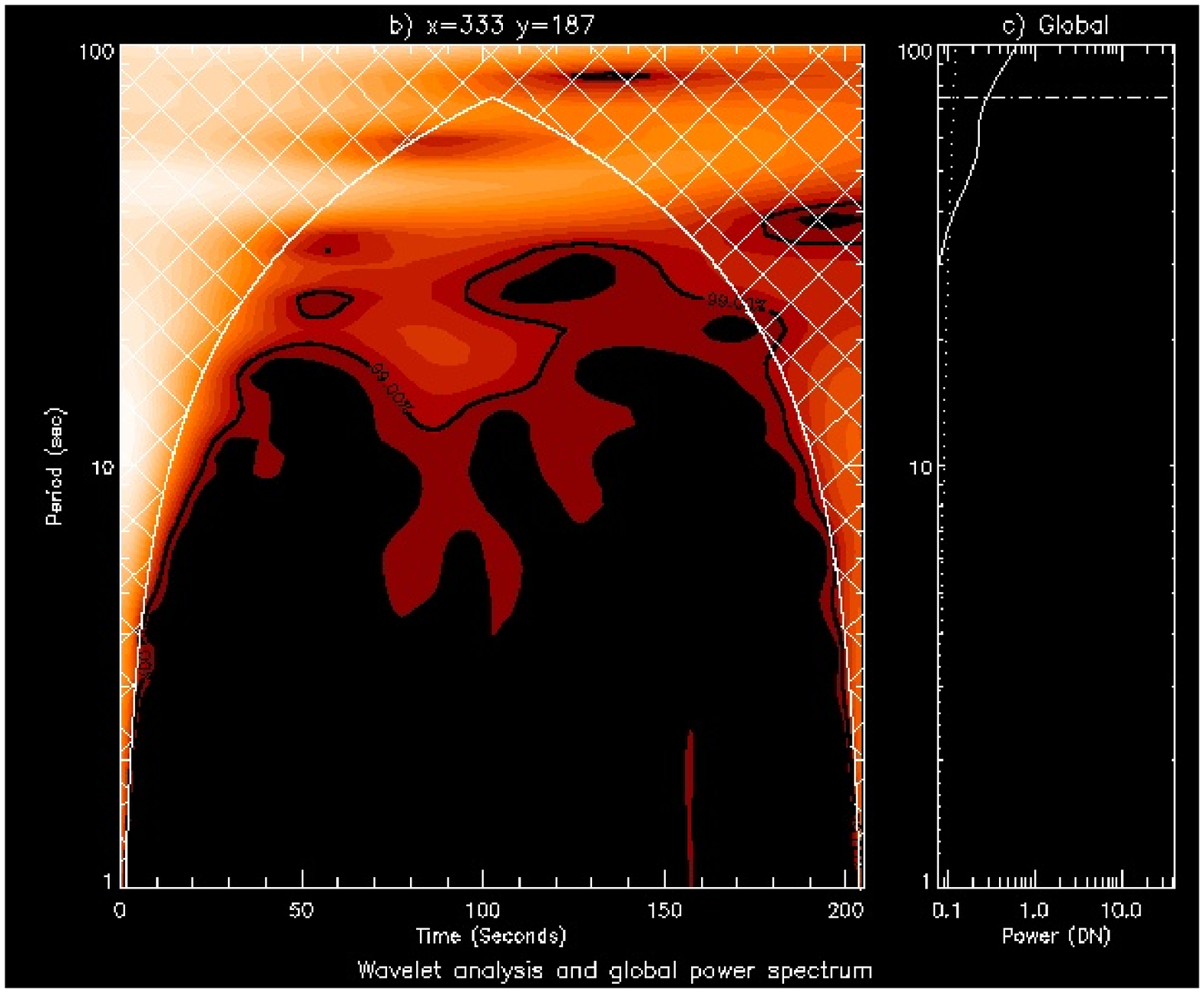}
\caption{The wavelet analysis of the same point as in Figure 1 after 
being filtered through the \`a Trous algorithm.}
\label{de-noised wavelet}
\end{figure}

Figure 2 contains the results of the wavelet analysis of the x=333,
y=187 pixel after filtering by the \`a Trous transformation. This is
the same point analysed previously in Figure 1 and a comparison
between the two will reveal the effect of the \`a Trous de-noising to
the data set. The filtering has affected the power spectrum in two
distinct ways, both of which can be associated with noise reduction.

Firstly, the lower the periodicity, the greater the reduction in
power. For our example, practically all oscillations below 4 s have
disappeared while those above 30~s are largely unaffected. This is
compatible with noise reduction since it is well understood that noise
tends to have a larger effect on frequencies that are closer to the
sampling rate (i.e., high frequencies are more influenced by noise).

Second, the oscillation at around $\sim$ 30 s that, based on previous
criteria (see [17] and discussion in next section), should be
considered \lq\lq real'' (i.e., caused by intensity oscillations in
the solar corona), is affected as expected. In particular, the
duration of the oscillation remains practically unchanged (from the
edge of the COI to 123$^{\rm rd}$ second), while the span of the
oscillation in the period axis has been reduced. This is again
compatible with noise reduction as detections caused by noise tend to
be very \lq\lq elongated'' on the period axis (i.e., have short
duration but stretch along many periodicities).

Interestingly, the shape of the oscillation in the wavelet space after
de-noising, bears a close resemblance to the theoretical prediction by
[5]. They produced the wavelet analysis of a numerically created,
impulsively, short-period magnetoacoustic wave train as it would had
been observed while propagating at distance {\it h} from its
source. Similarly to [5] simulations, the wavelet transformation of
Figure 2 contains a narrow spectrum tail leading to a much broader
band head (a \lq\lq tadpole'').

\section{\bf Automated Detection of Oscillations}

The main purpose of the SECIS observations is to detect coronal
intensity oscillations. Although S/N limitations confine the search of
such perturbations to the lower corona and more specifically in the
areas around coronal loops, these areas are sufficiently large to
require the development of software techniques to automatically detect
such events throughout the whole data set. Additionally, the same piece
of software will be used to deal with potential instrumental and
atmospheric effects (see Section 5).

For consistency with previous work (mainly [16] and [17]) we
implemented the following criteria for distinguishing between those
detections that are due to noise and those that are caused by signal
variations (either as a result of instrumental effects, the
atmosphere, or genuine coronal intensity variations):

\begin{itemize}

\item All coefficients falling within the COI are discarded.

\item Only those areas of a 99\% confidence level or higher are taken 
into account.

\item Oscillations lasting less than the time length of three periods are 
considered noise.

\end{itemize}

For a detailed discussion on the significance of these criteria, see 
[17]. A step-by-step description of the algorithm follows:

\begin{enumerate}

\item For a given pixel of the data set the 2-d wavelet coefficiencies 
of the time series is produced (as for Figures 2 and 3).

\item For the lowest periodicity of the analysis, the number of the 
first sample in time that is unaffected by the COI is determined. We
call this sample ${\rm t}$.

\item For the same periodicity the last sample that is unaffected by 
the COI, ${\rm t'}$, was determined.

\item The number of the sample that predates ${\rm t'}$ by three 
periodicities is determined. We call this sample ${\rm t}_{\mathrm
max}$

\item The confidence level of the sample ${\rm t}$ is extracted. 

\item If the confidence level is 99\% or higher, then the confidence
level of the sample that is three periods later than ${\rm t}$,
referred to here as ${\rm t_{+3}}$, is also extracted. Otherwise we
move to step 8.

\item If the confidence level of ${\rm t_{+3}}$ is also 99\% or higher
the co-ordinates of the pixel are recorded together with the current
periodicity and ${\rm t}$. In this case the algorithm moves to step
no.\ 9. It is assumed that if ${\rm t}$ and ${\rm t_{+3}}$ have both
confidence level of 99\% or higher, all samples between them will have
confidence level of 99\% or higher in the same periodicity.

\item Move to the next ${\rm t}$ and repeat steps 5-7 until ${\rm t}$ 
becomes ${\rm t_max}$.

\item Move to next periodicity and repeat steps 2-8 until the 
periodicity reaches the limit of 70.9 s.

\item Move to the next pixel of the array and start again from the 
beginning.

\end{enumerate}

For periods of more than 70.9 s the part of the time series that is
outside the COI is not longer than three periodicities.  For such
periods, the criteria established at the beginning of this section
cannot therefore be satisfied.

\section{\bf Instrumental and Atmospheric Effects}

The field-of-view covered by the SECIS observations can be roughly
divided into two areas. The first is covered by the lunar disk and
contains no signal apart from that of scattered light from the
atmosphere. The second area contains part of the lower corona and the
signal is dominated by light from the Sun's atmosphere. The third part
covers the outer corona and although most of the signal is again of
solar origin, a significant portion of it is due to earth's
atmosphere.

The automated method described in the previous sector was applied
first to the area covered by the moon disk in order to asses the
existence of instrumental and atmospheric oscillations. The signal
recorded from that area can only be from the scattered light of the
atmosphere and the CCD read-out noise, making it the most suitable
tool for determining bogus oscillations in the data set. By analysing
a rectangular area of 200$\times$50 pixel$^{2}$,~$\sim$2000 detections
of oscillations were made across the whole spectrum of
periodicities. We applied the same procedure in parts of the lower
(covering an area of 50$\times$250 pixel$^{2}$) and outer
(50$\times$300 pixel$^{2}$) corona and found $\sim$750 and $\sim$4000
oscillations respectively. All three areas are a minimum of 100 pixels
away from the left and right sides and 200 pixels away from the top
and bottom of the image. This is because instrumental oscillations are
known to appear in the edges of the CCD due to manufacturing
limitations. Also, detections that started during the first 1000
frames as well as those that finished during the last 1000 frames were
discarded as they may have been affected by light from the photosphere
during the start or end of the eclipse (an effect also known as \lq\lq
diamond rings effect'').

From the analysis of the above results it is apparent that a circular
area centred at the bottom part of the edge of moon's disk has a very
limited number of oscillations. A closer inspection revealed that this
area of the CCD has saturated by photospheric diamond rings. The rest
of the data set contains oscillations that are randomly scattered in
space in all periodicities.

\section{\bf Oscillation Detections}

\begin{figure}
\centering
\includegraphics[bb=36 184 560 659, width=8cm]{loop.eps}
\caption{Solar coronal loop of AR 9513 as observed by SECIS during 
the 2001 solar eclipse. The average intensity of the pixels is
displayed with the red-scale. Marked with crosses are the pixel
detected to oscillate in intensity according the criteria of section
4. }
\label{loop}
\end{figure}

From the areas of the lower corona analysed so far, the most
interesting is the loop of AR 9513 displayed in red-scale in Figure
3. Marked with crosses are the pixels that have been detected with
intensity oscillations with periods in the range of 20-30 s, while the
units of the x and y axes are pixel co-ordinates of the aligned data
set.  With the existence of atmospheric and instrumental detections in
the data set well established, it can be estimated that approximately
two thirds of the detections of Figure 3 are likely to be of solar
origin. This is for three main reasons:

\begin{itemize}

\item Compared to the nearby areas of the lunar disk and the outer 
corona which were scanned for oscillations, the area of Figure 3
contains $\sim$3 and 2 times more detections respectively. Whereas the
number of oscillations per unit spatial area remains constant,
regardless of the arbitrary area chosen, either on the lunar disk or
in the outer corona, this is not the case for the area of lower corona
presented in Figure 3.

\item The spatial distribution of the oscillations is non-random. 
Again, this is distinct from all areas of the moon and the outer
corona, with the only exception of the part of the data set affected
by the instrumental limitations described above, have a random scatter
of detections throughout the field.

\item The detections of Figure 3 form a shape that approximately 
\lq\lq coincides'' with a coronal loop of AR 9513.

\end{itemize}

The results described above are compatible with those based on the
SECIS 1999 observations as described by [15], [16], [17]. As seen in
Figure 3, there are two important confirmations of previously
published results. Firstly, there are intensity perturbations both
inside and exactly outside visible coronal loops. This is in line with
the findings of [15], [16] of oscillations inside and [17] of
oscillations outside coronal loops. Secondly, as previously reported
by [17], more intensity variations can be found outside the loops,
towards the tenuous part of the corona, than towards the denser part
(i.e., more oscillations can be detected at the higher altitudes of
the solar corona than the lower). For a more detailed discussion on
this result see [17].

The results presented above confirm the SECIS ability to detect MHD
oscillations and their presence in the data set of the 2001
observations. A detailed study of the detections reported above is to
follow and more work will be done in extending the search for
detections to other loops of AR 9513. Satellite observations of the
same active region shortly before, during and just after the eclipse
can be combined to determine various physical parameters useful in the
interpretation of the SECIS 2001 observations.

\section*{\bf Acknowledgments}

The authors would like to thank K.J.H. Phillips for his collaboration
on the SECIS project. PPARC funding was used during this work. ACK
acknowledges the Leverhume Trust for funding via grant F00203/A. JMA
\& DRW thank DEL and QUB for studentships. FPK is grateful to AWE
Aldermaston for the award of the William Penney fellowship.

\section*{\bf References}

\noindent 1.\ Aschwanden M. J., Fletcher L., Schrijver C. J., Alexander D., ApJ
520, 880, 1999

\noindent 2.\ Nakariakov V.M., in {\it Dynamic Sun}, Ed. B.~Dwivedi, CUP, 2003 

\noindent 3.\ Roberts B., in Proc., 10th European Solar Physics Meeting, {\it 
'Solar Variability: From Core to Outer Frontiers'}, ESA SP-506, 481, 2002 

\noindent 4.\ Roberts B., Edwin P.M., Benz A.O., 1984, ApJ, 279, 857

\noindent 5.\ Nakariakov V.M., Arber T.D., Ault C.E., Katsiyannis A.C., Williams 
D.R., 2004, MNRAS, submited

\noindent 6.\ Priest , E.R. \& Schrijver, C.J., 1999, Sol.\ Phys., 190, 1

\noindent 7.\ Parker, E.N., 1988, ApJ, 330, 474

\noindent 8.\ Hollweg, J.V., 1981, Sol.\ Phys., 70, 25

\noindent 9.\ Koutchmy, S., \v{Z}ug\v{z}da, Y.D. \& Loc\v{a}ns, V., 1983, A\&A, 
120, 185

\noindent 10.\ Pasachoff, J.M. \& Landman, D.A., 1984, Sol.\ Phys., 90, 325

\noindent 11.\ Singh, J., Cowsik, R., Raveendran, A.V., Bagare, S.P., Saxena, 
A. K., Sundararaman, K., Krishan, V., Naidu, N., Samson, J.P. A., Gabriel, F., 
1997, Sol.\ Phys., 170, 235

\noindent 12.\ Cowsik, R., Singh, J., Saxena, A.K., Srinivasan, R. \& Raveendran, 
A.V., 1999, Sol.\ Phys., 188, 89

\noindent 13.\ Pasachoff, J.M., Badcock, B.A., Russell, K.D. \& Seaton, D.B., 
2002, Sol.\ Phys., 207, 241

\noindent 14.\ Phillips, K.J.H., Read, P., Gallagher P.T., Keenan, F. P., 
Rudawy, P., Rompolt, B., Berlicki, A., Buczylko, A., Diego, F., Barnsley, R., 
Smartt, R.N., Pasachoff, J.M., Badcock, B.A., 2000, Sol.\ Phys., 193, 259.

\noindent 15.\ Williams, D.R., Phillips, K.J.H., Rudawy P., Mathioudakis, M., 
Gallagher, P. T., O'Shea, E., Keenan, F. P., Read, P., Rompolt, B., 2001, 
MNRAS, 326, 428

\noindent 16.\ Williams, D. R., Mathioudakis, M., Gallagher P.T., Phillips, 
K. J. H., McAteer, R. T. J., Keenan, F. P., Rudawy, P., Katsiyannis, A. C., 
2002, MNRAS, 336, 747

\noindent 17.\ Katsiyannis, A. C., Williams, D. R., McAteer, R. T. J., 
Gallagher, P. T., Keenan, F. P., Murtagh, F., 2003, A\&A, 406, 709

\noindent 18.\ Torrence \& C., Compo, G.P., 1998, Bull. Amer. Meteor. 
Soc., 79, 61

\noindent 19.\ Gallagher, P.T., Phillips, K.J.H., Harra-Murnion, L.K., 
Baudin, F. \& Keenan, F.P., 1999, A\&A, 348, 251

\noindent 20.\ Ireland, J., Walsh, R.W., Harrison, R.A. \& Priest, 
E.R., 1999, A\&A, 347, 355

\noindent 21.\ Banerjee, D., O'Shea, E. \& Doyle, J.G., 2000, A\&A, 355, 1152

\noindent 22.\ Starck, J.-L. \& Murtagh, F., 2002, `Astronomical Image and 
Data Analysis', Springer-Verlag Berlin Heidelberg


\end{document}